\documentclass[11pt,a4paper]{article}

\usepackage{amsmath, amssymb, graphics, graphicx,setspace,color}
\usepackage{multirow, array} 
\usepackage{ulem}
\newcommand{\be}{\begin{equation}}
\newcommand{\ee}{\end{equation}}
\newcommand{\ba}{\begin{eqnarray}}
\newcommand{\ea}{\end{eqnarray}}
\newcommand{\bas}{\begin{eqnarray*}}
\newcommand{\eas}{\end{eqnarray*}}

\newcommand{\bc}{\begin{center}}
\newcommand{\ec}{\end{center}}
\newcommand{\btab}{\begin{tabular}}
\newcommand{\etab}{\end{tabular}}

\newcommand{\nn}{\nonumber}

\newcommand{\mc}{\mathcal}

\newcommand{\al}{\alpha}
\newcommand{\bt}{\beta}
\newcommand{\g}{\gamma}         
\newcommand{\G}{\Gamma}
\newcommand{\dl}{\delta}        

\newcommand{\ep}{\epsilon}

\newcommand{\Tha}{\Theta}

\newcommand{\lm}{\lambda}

\newcommand{\pr}{\partial}

\newcommand{\nref}[1]{(\ref{#1})}

\begin{document}
\thispagestyle{empty}

\begin{flushright}LPT-Orsay-14-68\end{flushright}

\vspace{32pt}
\begin{center}
{\textbf{\Large On a boundary-localized Higgs boson in 5D theories}}

\vspace{40pt}

Roberto Barcel\'o~\footnote{roberto.barcelo@th.u-psud.fr}, Subhadip Mitra~\footnote{subhadip.mitra@th.u-psud.fr}, Gr\'egory Moreau~\footnote{gregory.moreau@th.u-psud.fr}
\vspace{12pt}

\textit{
Laboratoire de Physique Th\'eorique, Universit\'e Paris-sud 11 et CNRS-UMR 8627, \\
F-91405 Orsay Cedex, France.}\\ 
\end{center}

\vspace{40pt}

\begin{abstract}

In the context of a simple five-dimensional (5D) model with bulk matter coupled to a brane-localized Higgs boson, we point out a new non-commutativity in the 4D calculation of the mass spectrum for excited 
fermion towers~: the obtained expression depends on the choice in ordering the limits, $N \to \infty$ (infinite Kaluza-Klein tower) and $\epsilon \to 0$ ($\epsilon$ being the parameter introduced for regularizing 
the Higgs Dirac peak). This introduces the physical question of which one is the correct order~; we then show that the two possible orders of regularization (called I and II) are physically equivalent, as both 
can typically reproduce the measured observables, but that the one with less degrees of freedom (I) could be uniquely excluded by future experimental constraints. This conclusion is based on the exact 
matching between the 4D and 5D analytical calculations of the mass spectrum -- via the regularizations of type I and II. Beyond a deeper insight into the Higgs peak regularizations, this matching also allows 
us to confirm the validity of the usual 5D mixed-formalism and to clarify the UV cut-off procedure. All the conclusions, deduced from regularizing the Higgs peak through a brane shift or a smoothed square 
profile, are expected to remain similar in realistic models with a warped extra-dimension.

\end{abstract}

\newpage

\setcounter{page}{1}


\section{Introduction}

The recent and historical discovery of a Higgs-like boson~\cite{Djouadi:2005gi} around $125$~GeV at the Large Hadron Collider (LHC)~\cite{Hdisco} of the CERN fulfills the last 
missing piece of the particle content of the Standard Model (SM). However, even with the discovery of the Brout-Englert-Higgs scalar field~\cite{Higgs}, the mechanism responsible 
for breaking the ElectroWeak (EW) symmetry is not fully understood~; there remain some questions unresolved like, for example, determining the range of validity of the SM. 
If the SM is valid all the way up to the Planck scale then one can wonder why the EW energy scale (close to the Higgs mass) is so much smaller than the Planck scale.
The famous Randall-Sundrum (RS) proposition of an higher-dimensional background with the Higgs boson localized on a TeV or Infra-Red (IR) brane~\cite{Randall:1999ee}, 
besides addressing the gauge hierarchy problem of Higgs mass corrections, provides an aesthetic interpretation of this apparent discrepancy between fundamental scales of nature~: 
the measured Planck scale would be an effective four-dimensional (4D) scale whereas the gravity scale on the TeV-brane would be reduced by a warp factor down to the EW scale 
order. The later RS version with SM fields propagating in the bulk~\cite{Gherghetta:2000qt} even allows to explain the strong hierarchies among fermion masses.
\\
At this special moment where the LHC is scrutinizing the Higgs boson properties~\cite{HFits,Fit-last} and exploring higher energy frontiers, it is crucial for the community to have a deep 
theoretical understanding of the RS paradigm, in order to develop carefully phenomenological tests of such a scenario. These tests of the RS model can make use of the more and more 
precise experimental measurements in the Higgs sector~\cite{Goertz:2011hj,RSggH5D,RSfusion,RSHiggs} or of possible direct signatures from Kaluza-Klein (KK) excitations at 
colliders~\cite{LHCgluonKK,RSAFB,VKKRS,RSVL} (see Ref.~\cite{RSphREV} for a review).

Now, from the theoretical point of view, it turns out that recently there has been a debate in the literature on RS frameworks~\cite{Malm:2013jia,Carena:2012fk}. A certain 
non-commutativity has appeared~: different results were obtained for Higgs production/decay processes when taking $\epsilon \to 0$ and then $N_{KK} \to 
\infty$~\cite{Casagrande:2010si} or the opposite order~\cite{Azatov:2010pf}. $N_{KK}$ is the number of exchanged excited states at the level of the loop amplitude.
$\epsilon$ is the infinitesimal parameter introduced to regularize the Dirac peak along the extra-dimension associated to the Higgs scalar stuck on the IR-brane~; indeed, this Higgs 
peak induces the so-called jump problem, for the wave functions of the fermion bulk fields, which must be regularized. It was clearly crucial for testing the Higgs sector of the 
RS model at LHC to shed light on those theoretical subtleties.

In this paper, we show that there exists another type of non-commutativity in a 4D calculation (based on considering gradually KK tower effects)~: the fermion mass spectrum expression 
relies on the arbitrary choice in ordering the limits $\epsilon \to 0$ and $N \to \infty$, where $N$ is now the KK-index at the level of the calculation of mass eigenvalues.
We point out this non-commutativity in a toy model with a brane-localized Higgs boson and fermionic matter propagating along a flat extra-dimension, but our main 
conclusions are expected to remain true in more realistic warped extra-dimension scenarios. 
\\ 
So once more, it is urgent to really understand this new non-commutativity and to determine which order of the limits on $\epsilon$,$N$ has to be followed to construct a consistent 
model before studying its phenomenology. For that purpose, we perform calculations of the fermion mass spectrum, in both the 4D and 5D (based on equations of motion with Yukawa terms) 
approaches, applying consecutively the two possible orders -- assimilated to two kinds of Higgs regularization -- for the above mentioned limits on $\epsilon$,$N$. Those calculations 
allow effectively a better insight into the Higgs peak regularization features. 
\\
This 4D calculation of the mass spectrum reveals itself to be quite `heavy', due to the rich texture of the infinite fermion mass matrices, but it has the further interest to demonstrate 
analytically the exact matchings with the 5D calculation results. Obtaining these 4D/5D matching results represents the opportunity to confirm the 5D formalism for KK mixings  
used in literature and also to establish clearly a Ultra-Violet (UV) cut-off procedure in this context. 
\\ 
Let us finally specify that in order to provide various illustrations of our calculations within the above two types of regularizations, we regularize the Higgs delta peak by shifting it away 
from the boundary as well as smoothing it into a square profile -- which constitutes an equivalent alternative.

The paper is organized accordingly to the following simple plan. While the Section~2 is devoted to the 5D approach of the fermion mass spectrum, 
Section~3 is focused on the 4D treatment and the two calculations are compared in the synthesis made in Section~4. We summarize and conclude in Section~5.


\section{5D calculations}\label{sec:5d}

\subsection{The model}\label{sec:themodel}

We consider a toy model with an extra spatial dimension having a flat geometry and being parametrized by the coordinate, $y$. 
This extra-dimension constitutes an interval of length $\pi R$ with two boundaries at $y=0,\pi R$. The Higgs boson of the SM, embedded in a doublet 
under the $SU(2)_L$ gauge group, is strictly localized on the brane at $y=\pi R$ while some fermionic matter is spread out in the bulk.
For illustration, let us consider the first quark generation~\footnote{The third family of heavy SM quarks is generally expected to feel the largest mixings 
with KK modes, but our formalism is directly extendable to any quark generation as well as to leptons.}~;
the down-quark fields denoted by $Q$ and $D$ are respectively a doublet component and a singlet under $SU(2)_L$, as in the SM. 
The dynamics for the up-quark sector fields, $\tilde Q$ and $U$, is dictated by an identical Lagrangian and thus we will not repeat such
an analog analysis throughout the paper. For our task, it is sufficient to concentrate on the kinetic terms for the down-quarks as well as their 
Yukawa interactions, whose fundamental 5D action can be written as usual (after the EW symmetry breaking),
\ba
S_{\textrm{fermion}}&=&\int \textrm{d}^4x~\textrm{d}y~\big{\lbrack}~ \frac{i}{2}(\bar{Q}\Gamma^M \pr_M Q ~-~\pr_M \bar{Q} \Gamma^M Q ~+ \{ Q \leftrightarrow D \} )\nonumber\\
~&&-~ \delta(y-\pi R) \ (Y_5\ \bar{Q}_LHD_R ~+~ Y^\prime_5\ \bar{Q}_RHD_L ~+~  \textrm{H.c.})~ \big{\rbrack}\;,\label{eq:action}
\ea
where the index is $M=0,1,2,3,5$ and the Higgs field is developed into the 4D scalar plus its vacuum expectation value as, 
$H=\frac{v+h(x)}{\sqrt{2}}$, $x$ representing the usual four coordinates.
It should be remarked that the coupling constants $Y_5$ and $Y^\prime_5$ are independent~; in order to avoid the introduction of a new scale in the theory, 
one can choose $Y_5=yR$ and $Y^\prime_5=y'R$, where $y,y'$ are dimensionless coupling constants of ${\cal O}(1)$. In our notations, the 5D Dirac spinor, 
being the smallest irreducible representation of the Lorentz group, reads as,
\begin{equation}
Q = \left(\begin{array}{c} Q_L \\ Q_R   \end{array}\right) \;\;\;\; \mbox{and} \;\;\;\;   D = \left(\begin{array}{c} D_L \\ D_R   \end{array}\right)\;,
\label{5DDiracSp}
\end{equation}
in terms of the two two-component spinors, for the two down-quark fields.

\subsection{The KK decomposition and equations of motion}\label{sec:theKKdecomp5D}

In this Section~\ref{sec:5d}, we derive the fermion masses using the so-called exact or 5D approach. In this approach, one keeps the Yukawa mass terms that appear after EW
symmetry breaking in the equations of motion for the fermion profiles along the extra-dimension [we will simply refer to those as Equations Of Motion (EOM)]. The advantage 
of this approach is that the mixing among {\it all} KK modes for any fermion is automatically taken care of when solving the EOM's. Hence this method for deriving the masses is 
called a 5D calculation as it incorporates the full effect of the 5D theory in the EOM's.

The first step is to perform a `mixed' KK decomposition of the 5D fields in Eq.~(\ref{5DDiracSp}),
\begin{subequations}\begin{align}
Q_L &= \sum_{n=0}^{\infty} q_L^n(y)\;Q_L^n (x)\;, \label{first5Dfield} \\
Q_R &= \sum_{n=0}^{\infty} q_R^n(y)\;D_R^n (x)\;, \\
D_L &= \sum_{n=0}^{\infty} d_L^n(y)\;Q_L^n (x)\;, \\
D_R &= \sum_{n=0}^{\infty} d_R^n(y)\;D_R^n (x)\;, \label{last5Dfield}
\end{align}\end{subequations}
where $Q_L^n(x)$, $D_R^n(x)$ are the 4D fields 
and $q_{L,R}^n(y)$, $d_{L,R}^n(y)$ are the corresponding wave functions along the extra-dimension. 
Although not essential for our calculations, we note for completeness that with this KK decomposition, the profiles satisfy the following normalization condition,
\ba
\int_0^{\pi R} dy  \left[\vert q_X(y)\vert^2+\vert d_X(y)\vert^2\right] =1\;; \ \  \mbox{with } X=L,R\;.\nonumber
\ea
Through a factorization of the 4D fields, the mixed KK decomposition allows to separate the Euler-Lagrange equations for the 5D fields into the 4D Dirac equations
($\mu=0,1,2,3$),
\ba
-i \bar{\sigma}^{\mu} \partial_{\mu} Q^n_L(x) + m~D^n_R(x) &=& 0\;, \\
-i \sigma^{\mu} \partial_{\mu} D^n_R(x) + m~Q^n_L(x) &=& 0\;,
\ea
and the equations of motion for any excited fermion profile after EW symmetry breaking, 
\begin{subequations}\begin{align}
-\ m~q_L ~-~ q'_R ~+~ \dl (y-\pi R) \ \frac{v Y_5}{\sqrt{2}}~d_R ~&=~ 0\;, \label{ppprofileD} \\
-\ m~q_R ~+~ q'_L ~+~ \dl (y-\pi R) \ \frac{v Y^\prime_5}{\sqrt{2}}~d_L ~&=~ 0\;, \label{mmprofileD} \\
-\ m~d_L ~-~ d'_R ~+~ \dl (y-\pi R) \ \frac{v Y^\prime_5}{\sqrt{2}}~q_R ~&=~ 0\;, \label{mmprofileQ} \\
-\ m~d_R ~+~ d'_L ~+~ \dl (y-\pi R) \ \frac{v Y_5}{\sqrt{2}}~q_L ~&=~ 0\;, \label{ppprofileQ} 
\end{align}\end{subequations}
where the `~$'$~' exponent after any wave function denotes the derivative with respect to the fifth coordinate, $y$. 
We have assumed real Yukawa coupling constants and $m$ masses for simplicity, but this kind of analysis is generalizable to cases with complex phases.

The variation of the action combined with the above EOM's on the boundaries give rise either to the Dirichlet Boundary Conditions (BC) on both boundaries [i.e. vanishing profiles
at the two endpoints], denoted $(--)$ and to be assigned to $q_R$ and $d_L$, or to the Neumann BC [vanishing derivatives], noted $(++)$ and assigned to $q_L$ and $d_R$.
Now, due to the $\dl (y-\pi R)$-term in Eq.~(\ref{ppprofileD}), its infinitesimal integration around $y=\pi R$ leads to two distinct values of $q_R$ at that point, 
which together with the unique $q_R$ $(--)$ BC renders the value of this profile at $y=\pi R$ ambiguous~: this is the `jump' problem~\cite{Azatov:2009na}, first described 
on an interval in Ref.~\cite{Csaki:2003sh}, which also arises of course for the $d_L$ profile in Eq.~(\ref{ppprofileQ}).
\\
To avoid this ambiguity one has to regularize the Higgs peak~\cite{Csaki:2003sh}~: this can be done via shifting the Dirac peak away from the boundary by a small ($\ep R$) amount, 
or, via smoothing the peak by giving it a narrow width (like a normalized square function of width $\ep R$). Then one imposes the $(--)$ BC's and solves the EOM's (involving $\ep$) 
to find the fermion masses, before finally taking the limit, $\ep\to0$, in order to recover the wanted brane-localized Higgs situation. We are going to realize explicitly those two schemes of 
$\ep$-regularization in the next two subsections.

\subsection{Moving the Higgs peak}\label{sec:5d_deltahiggs}

If one shifts the Higgs peak by a distance $\ep R$ away from the $\pi R$-boundary, 
\ba
 \dl \left(y-\pi R\right)\rightarrow  \dl \left(y-(\pi-\ep)R\right)\;,\label{deltashift}
\ea
then profile jumps move from the boundary to the bulk. The EOM's that one needs to solve become,  
\begin{subequations}\begin{align}
-\ m~q_L ~-~ q'_R ~+~ \dl \left(y-(\pi-\ep)R\right) \ \frac{vY_5}{\sqrt2}~d_R = 0\;, \label{eq:5deom_delta_a}\\
-\ m~q_R ~+~ q'_L ~+~ \dl \left(y-(\pi-\ep)R\right) \ \frac{vY^\prime_5}{\sqrt2}~d_L = 0\; ,\label{eq:5deom_delta_b}\\
-\ m~d_L ~-~ d'_R ~+~ \dl \left(y-(\pi-\ep)R\right) \ \frac{vY^\prime_5}{\sqrt2}~q_R = 0\;, \label{eq:5deom_delta_c}\\
-\ m~d_R ~+~ d'_L ~+~ \dl \left(y-(\pi-\ep)R\right) \ \frac{vY_5}{\sqrt2}~q_L = 0\;. \label{eq:5deom_delta_d}
\end{align}\end{subequations}
Solving this set of equations is not very complicated since, for $0\leq y < (\pi-\ep)R$ and $(\pi-\ep)R < y \leq \pi R$, the above equations become identical to the free equations, 
i.e. EOM's without the Yukawa terms. For the $q_R$, $d_L$ solutions satisfying the $(--)$ BC and the $q_L$, $d_R$ solutions with $(++)$ BC, at $y=0$
[BC's still induced by the action variation combined with the new EOM's~(\ref{eq:5deom_delta_a})-(\ref{eq:5deom_delta_d}) on the boundaries], 
we get the following physical profiles,
\ba
q_L(y)  = \mc C \cos(m y)\;,\quad q_R(y)  =-\mc C \sin(m y)\;, \quad
 d_R(y) = \mc C \cos(m y)\;,\quad d_L(y)  =\mc C  \sin(m y)\;,\label{eq:free_5d_sols}
\ea
which are valid for $0\leq y < (\pi-\ep)R$. $\mc C$ denotes the normalization factor. For $(\pi-\ep)R < y \leq \pi R$, we obtain the following general EOM solutions,
\ba
&&\hat q_L(y)  = B_1 \cos(m y) + B_2 \sin(m y),\quad \hat q_R(y)  = B_2 \cos(m y) - B_1 \sin(m y), \nn\\
&&\hat d_L(y)  = B_3 \cos(m y) + B_4 \sin(m y), \quad \hat d_R(y)  = B_4 \cos(m y) - B_3 \sin(m y), \label{eq:free_5d_solsII}
\ea
where $B_i$'s are arbitrary constants that are fixed by the normalizations. From Eq.~\nref{eq:5deom_delta_a}-\nref{eq:5deom_delta_d}, we see that the amount of jump that a field 
undergoes is proportional to the value of some other profile exactly at $y=(\pi-\ep)R$. Hence to connect the two sets of solutions in Eq.~\nref{eq:free_5d_sols} and
Eq.~\nref{eq:free_5d_solsII}, one needs to assign some values for these profiles at the jump point. We use the following convention for a generic profile,
\ba
f\left((\pi-\ep)R\right) = \frac1{1+c}\left[f\left((\pi-\ep)R\right)+c \hat f\left((\pi-\ep)R\right)\right],
\label{c=1prescription}
\ea
i.e. we take the weighted average of the limiting values of the function approaching from both sides, which, for $c=1$, translates into the normal averaging. The continuity
conditions read then as,
\begin{subequations}\begin{align}
q_R\left((\pi-\ep)R\right) - \hat q_R\left((\pi-\ep)R\right) =& \frac{-vY_5}{\sqrt2 (1+c)} \big[d_R\left((\pi-\ep)R\right) + c\, \hat d_R\left((\pi-\ep)R\right)\big]\;,\\
q_L\left((\pi-\ep)R\right) - \hat q_L\left((\pi-\ep)R\right) =& \frac{vY^\prime_5}{\sqrt2 (1+c)} \big[d_L\left((\pi-\ep)R\right) + c\, \hat d_L\left((\pi-\ep)R\right)\big]\;,\\
d_R\left((\pi-\ep)R\right) - \hat d_R\left((\pi-\ep)R\right) =& \frac{-vY^\prime_5}{\sqrt2 (1+c)} \big[q_R\left((\pi-\ep)R\right) + c\, \hat q_R\left((\pi-\ep)R\right)\big]\;,\\
d_L\left((\pi-\ep)R\right) - \hat d_L\left((\pi-\ep)R\right) =& \frac{vY_5}{\sqrt2 (1+c)} \big[q_L\left((\pi-\ep)R\right) + c\, \hat q_L\left((\pi-\ep)R\right)\big]\;.
\end{align}\end{subequations}
Injecting Eq.~\nref{eq:free_5d_sols}-\nref{eq:free_5d_solsII} in these four relations gives us the following constant expressions,
\ba
B_1 = B_4 &=& \frac{\mc C [(1 + c)^2 (X+X^\prime) \sin\left(2  (\pi  - \ep) mR \right) -2( (1+ c)^2 + c XX^\prime)] }{2[c^2XX^\prime-(1+c^2)]}\;,\\
B_3 = -B_2 &=& \frac{\mc C  (1 + c)^2[X-X^\prime+ (X+X^\prime) \cos\left(2  (\pi  - \ep) mR\right) ] }{2[c^2XX^\prime-(1+c^2)]}\;,
\ea
where $X=vY_5/\sqrt2$ and $X^\prime = vY^\prime_5/\sqrt2$. One can now apply the BC for the $(--)$ modes on the $y=\pi R$ brane,
\ba
\hat q_R(\pi R) = \hat d_L(\pi R) =0\;.
\ea
For small $\ep\to0$, this requires,
\ba
\tan\left(\pi R\; m\right) = \frac{\sqrt2(1+c)^2 vY_5}{2(1+c)^2+cv^2Y_5Y^\prime_5}\;,\label{eq:5ddeltahiggspres}
\ea
which for $c=1$ becomes,
\ba
\tan\left(\pi R\; m\right) &=& \frac{4\sqrt2 vY_5}{8+v^2Y_5Y^\prime_5}\;.\label{eq:5ddeltahiggs}
\ea
This relation gives directly the solutions for the fermion mass spectrum.

It is possible to choose another order of calculation. Indeed, one can first apply the BC for the $(--)$ modes on the $y=\pi R$ brane,  
so that the Yukawa terms in the EOM's~\nref{mmprofileD}-\nref{mmprofileQ} vanish, and then solve these  
EOM's with a $\ep R$-shifted peak. In the absence of such Yukawa terms, the profiles for the $(++)$ modes become continuous across 
$y=(\pi-\ep)R$ and the new continuity conditions are,
\begin{subequations}\begin{align}
q_R\left((\pi-\ep)R\right) - \hat q_R\left((\pi-\ep)R\right) \ =&\  \frac{-vY_5}{\sqrt2 }  d_R\left((\pi-\ep)R\right)\;,\\
q_L\left((\pi-\ep)R\right) - \hat q_L\left((\pi-\ep)R\right) \ =&\  0\;,\\
d_R\left((\pi-\ep)R\right) - \hat d_R\left((\pi-\ep)R\right) \ =&\  0\;,\\
d_L\left((\pi-\ep)R\right) - \hat d_L\left((\pi-\ep)R\right) \ =&\  \frac{vY_5}{\sqrt2}q_L\left((\pi-\ep)R\right)\;.
\end{align}\end{subequations}
There is thus no need to choose any prescription like in Eq.~\nref{c=1prescription}. Following the same steps of calculation as before --
except this difference on the absence of Yukawa terms --
we arrive at the following simpler result for the fermion mass spectrum, dictated by the $(--)$ BC at $y=\pi R$, in the limit $\ep\to0$,
\ba
\tan\left(\pi R \;m\right) &=& \frac{vY_5}{\sqrt2}\;.\label{deltahiggs5dpp}
\ea
This mass result can also be obtained from Eq.~\nref{eq:5ddeltahiggspres} by setting $Y^\prime_5=0$.

\subsection{Smoothing the Higgs peak}\label{5Dbox}

We can alternatively replace the Higgs Dirac peak at the boundary by a normalized square function, of width $\ep R$ and height $1/\ep R$, so that the Dirac peak is recovered in the
limit, $\ep \to 0$. With this smooth profile, one gets the following EOM's,
\begin{subequations}\begin{align}
-\ m~q_L ~-~ q'_R ~+~ \frac{\Tha\left(y-(\pi-\ep)R\right)}{\ep R} \ \frac{vY_5}{\sqrt2}~d_R = 0\;, \label{eq:5deom_sq_a}\\
-\ m~q_R ~+~ q'_L ~+~ \frac{\Tha\left(y-(\pi-\ep)R\right)}{\ep R} \ \frac{vY^\prime_5}{\sqrt2 } d_L = 0\;,\label{eq:5deom_sq_b}\\
-\ m~d_L ~-~ d'_R ~+~ \frac{\Tha\left(y-(\pi-\ep)R\right)}{\ep R} \ \frac{vY^\prime_5}{\sqrt2 }~q_R = 0\;, \label{eq:5deom_sq_c}\\
-\ m~d_R ~+~ d'_L ~+~ \frac{\Tha\left(y-(\pi-\ep)R\right)}{\ep R} \ \frac{vY_5}{\sqrt2 }~q_L = 0\;, \label{eq:5deom_sq_d}
\end{align}\end{subequations}
where $\Tha(y) =1$ for $y\geq0$ and zero otherwise. In the range $0\leq y < (\pi-\ep)R$, these equations correspond to the free EOM's and have the same solutions as in 
Eq.~\nref{eq:free_5d_sols} if we impose once more the $(--)$ and $(++)$ BC's at $y=0$. Assuming $Y_5=Y^\prime_5$ for simplicity, the following generic 
ansatz solves the EOM's~\nref{eq:5deom_sq_a}-\nref{eq:5deom_sq_d} in the range $(\pi-\ep)R \leq y \leq \pi R $ ,
\ba
\hat f_X(y) &=& A_{f_X} \exp\left(\sqrt{\frac{v^2 Y_5^2 - 2 m^2 \ep^2 R^2}{2 \ep^2 R^2}}y\right) + B_{f_X} \exp\left(-\sqrt{\frac{v^2 Y_5^2 - 2 m^2 \ep^2 R^2}{2 \ep^2 R^2}}y\right)\;, 
\ea
$f_X$ standing for any profile and $A_{f_X}$,$B_{f_X}$ being normalization constants.
Demanding that all the profiles are continuous across $y=\left(\pi-\ep\right)R$ and setting $\hat q_R(\pi R) = \hat d_L(\pi R) = 0$ 
[BC for the $(--)$ modes] gives us the following condition on the mass,
\ba
\tan\left( \pi R \; m\right) &=& \sqrt{\frac{vY_5 - \sqrt2m\ep R}{vY_5 + \sqrt2m\ep R}} \;\tanh\left(\sqrt{\frac{v^2Y_5^2 - 2m^2\ep^2  R^2}{2}}\right)\;.
\ea
In the limit $\ep\to0$ this simplifies to,
\ba
\tan\left(\pi R \;m\right) &=& \tanh\left(\sqrt{\frac{v^2Y_5^2}{2}}\right)\;.
\label{SpecBox5DII}
\ea

As in the case of the shifted delta function, if one first imposes instead the BC for the $(--)$ modes, $\hat q_R(\pi R) = \hat d_L(\pi R) =0$,   
the Yukawa terms in the Eq.~\nref{mmprofileD}-\nref{mmprofileQ} are eliminated. Then solving the   
EOM's~\nref{eq:5deom_sq_a}-\nref{eq:5deom_sq_d} with an $\ep R$-square function but without those two Yukawa terms, one recovers,
through the same steps of calculation, the simple mass spectrum of Eq.~\nref{deltahiggs5dpp}.


\section{4D calculations}\label{sec:4d}

\subsection{The KK decomposition and mass matrices}\label{sec:theKKdecomp4D}

In this Section~\ref{sec:4d}, considering the same model as the one defined by the Lagrangian~\nref{eq:action}, 
we calculate the fermion masses in the maybe more intuitive approach referred to as the perturbative or 4D calculation. 
To obtain the fermion profiles, here, one considers the free EOM's, i.e. the equations without Yukawa mass terms. As a result, unlike the 5D point of view  
addressed in the previous Section~\ref{sec:5d}, one needs to diagonalize the fermion mass matrices to include the whole KK mass
mixing effect. The 4D approach denomination relies on the fact that one starts from a 4D model without KK modes and the entire KK tower is taken into account gradually,  
through the limit $N\to \infty$. It is also called a perturbative approach in the sense that the Yukawa interaction is incorporated via infinite series terms. 
\\
Now, these infinite numbers of KK excitations lead to mass matrices of infinite dimensions whose exact diagonalization can represent a challenging task. However, in certain 
cases it is possible analytically as we shall illustrate in this Section~\ref{sec:4d}. The aim being to compare the fermion masses obtained by diagonalizing the complete 
mass matrix with the ones obtained from the previous 5D approach.

We start by decomposing the 5D fields in KK towers like,
\begin{subequations}\begin{align}
Q_L &= \sum_{n=0}^{\infty} q_L^n(y)\;Q_L^n (x)\;, \label{Abasic5Dfield} \\
Q_R &= \sum_{n=0}^{\infty} q_R^n(y)\;Q_R^n (x)\;, \\
D_L &= \sum_{n=0}^{\infty} d_L^n(y)\;D_L^n (x)\;, \\
D_R &= \sum_{n=0}^{\infty} d_R^n(y)\;D_R^n (x)\;, \label{Dbasic5Dfield} 
\end{align}\end{subequations}
which gives rise to the following KK mass terms in the 4D effective Lagrangian,
\ba
\mc L_{\rm KK} = -\sum_{n=0}^\infty \left[M_{qn} \bar Q_L^n(x) Q_R^n(x) ~+~M_{dn} \bar D_L^n(x) D_R^n(x) \right] + {\rm H.c.}
\nonumber
\ea
where
\ba
M_{qn} = M_{dn} = \frac{n}{R}\;. \label{eq:Mqn}
\ea
The complete quark mass matrix in the 4D effective picture, after EW symmetry breaking, reads as,
\ba
\mc L_{\rm mass} = -\bar\Psi_L\,\cdot\,\left[M\right]\,\cdot\Psi_R + {\rm H.c.} \nonumber
\ea
and can be expressed, in the `combined' basis for the Left and Right-handed fields, 
\ba
\Psi^t_L &=& (Q_L^0,D_L^0,Q_L^1,D_L^1,Q_L^2,D_L^2,\cdots) \;, \nonumber \\ 
\Psi^t_R &=& (Q_R^0,D_R^0,Q_R^1,D_R^1,Q_R^2,D_R^2,\cdots)\;,\label{WeakBasis}
\ea
by the following infinite matrix,
\ba
\left[M\right]& \equiv &\left( \begin{array}{ccccccc}  
M_{q0} & \alpha_{00} & 0 & \alpha_{01 } & 0 & \alpha_{02} & \cdots \\
\beta_{00} & M_{d0} & \beta_{01} & 0 & \beta_{02} & 0 & \cdots \\
0 & \alpha_{10} & M_{q1} & \alpha_{11} & 0 & \alpha_{12} & \cdots \\
\beta_{10} & 0 & \beta_{11} & M_{d1} & \beta_{12} & 0 & \cdots\\
0 & \alpha_{20} & 0 & \alpha_{21} & M_{q2} &  \alpha_{22} & \cdots \\
\beta_{20} & 0 & \beta_{21} & 0 & \beta_{22} & M_{d2}  & \cdots \\
\vdots & \vdots & \vdots & \vdots & \vdots & \vdots & \ddots
\end{array} \right) ,
\label{eq:MassMat}
\ea
with,
\ba
\al_{ij} &=& Y_5\int_0^{\pi R} dy \ \delta(y-\pi R) \ \frac{v}{\sqrt 2} \ q_L^i (y) \ d_R^j(y)\;,\label{def:alpha}\\
\bt_{ij} &=& Y^\prime_5\int_0^{\pi R} dy \ \delta(y-\pi R) \ \frac{v}{\sqrt 2} \ d_L^i (y) \ q_R^j(y)\;.\label{def:beta}
\ea

To try to match the different regularizations performed in the 5D approach of Section~\ref{sec:5d}, we will treat similarly the Higgs peak -- by either moving or smoothing it -- 
in the 4D calculations of next two subsections.

\subsection{Moving the Higgs peak}\label{sec:4d_deltahiggs}

The fields~(\ref{WeakBasis}) undergo the unitary transformation matrices to the physical basis and the squared modulus of the quark masses, $\vert m \vert^2$, 
are the eigenvalues, noted $\lambda$, of the infinite-dimensional matrix, $\left[M^\dag M\right]$. For a general Higgs profile, we present in the Appendix  
one of the main results of the paper~: the Characteristic Equation (CE), for the infinite $\left[M^\dag M\right]$ matrix,
whose solutions are the eigenvalues, $\lambda = \vert m \vert^2$. From the obtained expression of the CE terms shown there, a logical structure in series 
emerges for such a general case. The CE contains infinite series of various types which can be written using the generic structures, $A_n$ and $B_n$, 
involving respectively, $\al_{ij}$ and $\bt_{ij}$. 

Let us now focus on the case of a Higgs peak infinitesimally shifted at some point, $y=(\pi-\ep) R$, along the extra-dimension as in Eq.~(\ref{deltashift}). 
Then the CE takes a much simpler form since the functions, $\al_{ij}$ and $\bt_{ij}$, are factorizable in, $i$ and $j$,
\ba
 \al_{ij}&=& \frac{vY_5}{\sqrt2}\, q_L^i ((\pi-\ep)R) \times d_R^j((\pi-\ep)R)\;, \label{def:alpha_dl}\\
 \bt_{ji}&=& \frac{vY^\prime_5}{\sqrt2}\, q_R^i ((\pi-\ep)R) \times d_L^j((\pi-\ep)R)\label{def:beta_dl}\;,
\ea
so that accordingly to Eq.~(\ref{ASnature}), 
\ba
A_{n>1} = B_{n>1} =0\ ,
\ea
due to the anti-symmetric constructions of $A_n$ and $B_n$. As a result the generic CE of Eq.~(\ref{GenCE}) simplifies to, 
\ba
&1 &+ \sum_{q_1;d_1}(-\lm)\frac{\left(\al_{q_1d_1}\right)^2+\left(\bt_{d_1q_1}\right)^2}{(M_{q_1}^2-\lm)(M_{d_1}^2-\lm)} +\sum_{q_1,q_2;d_1,d_2}(-\lm)^2
\frac{\left(\al_{q_1d_1}\right)^2\left(\bt_{d_2q_2}\right)^2}{(M_{q_1}^2-\lm)(M_{d_1}^2-\lm)(M_{q_2}^2-\lm)(M_{d_2}^2-\lm)}\nn\\
&
\times&\left(1-\dl_{q_1q_2}\frac{M_{q_2}^2}{\lm}\right)
\left(1-\dl_{d_1d_2}\frac{M_{d_2}^2}{\lm}\right) 
-\sum_{Q_1;D_1} 2 \ M_{Q_1}M_{D_1} \frac{\al_{Q_1D_1} \bt_{D_1Q_1}}{(M_{Q_1}^2-\lm)(M_{D_1}^2-\lm)}\nn \\
&+&\sum_{Q_1<Q_2:d_1,d_2}  \frac{2  (-\lm) M_{Q_1}M_{Q_2}}{(M_{Q_1}^2-\lm)(M_{Q_2}^2-\lm)} 
\times\frac{\al_{Q_1d_1}\al_{Q_2d_1} \bt_{d_2Q_1}\bt_{d_2Q_2}}{(M_{d_1}^2-\lm)(M_{d_2}^2-\lm)}\times\left(1-\dl_{d_1d_2}\frac{M_{d_2}^2}{\lm}\right)\nn\\
&+&\sum_{q_1,q_2:D_1<D_2}  \frac{2  (-\lm) M_{D_1}M_{D_2}}{(M_{D_1}^2-\lm)(M_{D_2}^2-\lm)} \times\frac{\al_{q_1D_1}\al_{q_1D_2}
\bt_{D_1q_2}\bt_{D_2q_2}}{(M_{q_1}^2-\lm)(M_{q_2}^2-\lm)}\times\left(1-\dl_{q_1q_2}\frac{M_{q_2}^2}{\lm}\right)\nn\\
&+&\sum_{Q_1<Q_2;D_1<D_2} 2\left(\prod_{i=1,2}\frac{ M_{Q_i}M_{D_i}}{(M_{Q_i}^2-\lm)(M_{D_i}^2-\lm)}\right) \nn\\
& \times & \Big( \al_{Q_1D_1}\al_{Q_2D_2}\times \bt_{D_1Q_1}\bt_{D_2Q_2} 
+  \al_{Q_1D_2}\al_{Q_2D_1}\times \bt_{D_2Q_1}\bt_{D_1Q_2}\Big) = 0\;. \label{eq:che_fac}
\ea
Here and elsewhere, unless specified, a sum over any index is assumed to be running from $0$ to $\infty$~; in the above relation, the KK masses obey
{\it e.g.}, $M_{q_1} = q_1/R$ where $q_1$ is a running integer [slightly different writing from Eq.~(\ref{eq:Mqn}) to ease notations]. 
We stress that to derive Eq.~\nref{eq:che_fac}, no approximation has been made, or in other words this equation exhibits the complete CE in this case.
Choosing the $(--)$ and $(++)$ BC's [from the action variation and free EOM's on boundaries] for the quark profiles, to end up with a chiral theory, 
we get the following normalized solutions of the free EOM's~\footnote{Although we have kept $\bt_{0j}$,$\bt_{i0}$ in the matrix~(\ref{eq:MassMat}) to make its 
($\alpha_{ij} \leftrightarrow \bt_{ji}$) symmetric texture explicit, we note that $\bt_{0j} = \bt_{i0} = 0$ since the zero-modes $q^0_R(y)=d^0_L(y) =0$.}, 
\ba
q^n_L(y) = d^n_R(y)=\sqrt{\frac{2}{\pi R}}\cos\left(\frac{ny}{R}\right) \;&,&\quad
-q^n_R(y) = d^n_L(y)=\sqrt{\frac{2}{\pi R}}\sin\left(\frac{ny}{R}\right)\; \quad \textrm{for} \;\; n >0 \nonumber \\ 
q^0_L(y) = d^0_R(y)=\sqrt{\frac{1}{\pi R}}\;&,&\quad  -q^0_R(y) = d^0_L(y)=0 \quad \quad \textrm{for} \;\; n =0 \ .
\label{eq:4dprofiles}
\ea 
With these solutions, the $\al_{ij}$ and $\bt_{ji}$ functions of Eq.~(\ref{def:alpha_dl})-(\ref{def:beta_dl}) become,
\ba
 \al_{ij}&=& \frac{\sqrt2 vY_5}{\pi R}\, \cos (i(\pi-\ep)) \cos(j(\pi-\ep))\;,\ \ \al_{00}=\frac{v Y_5}{\sqrt2 \pi R}\;,\label{def:alpha_dl_flat}\\
 \bt_{ji}&=& \frac{- \sqrt2 vY^\prime_5}{\pi R}\, \sin (i(\pi-\ep)) \sin(j(\pi-\ep))\label{def:beta_dl_flat}\;,\ \ \bt_{j0}=\bt_{0i}=\bt_{00}=0\;.
\ea
We are now in possession of all the necessary tools to simplify and solve the CE~\nref{eq:che_fac} in terms of the mass. 
Computing analytically all the involved infinite sums (over KK modes), we find the following compact form for the CE, in the final limit $\ep\to0$,
\ba
1 + \frac{1}{4} v^2Y_5Y^\prime_5 + \frac{1}{64} v^4(Y_5Y^\prime_5)^2 &=& \frac{v^2Y_5^2}{2} \cot^2{\left(\pi R \sqrt\lm\right)} \ , \label{deltahiggs4d1}\\
{\rm or,}\quad \tan^2\left(\pi R\sqrt{\vert m \vert^2}\right) &=&\left(\frac{4\sqrt2 vY_5}{8+v^2Y_5Y^\prime_5}\right)^2 \label{deltahiggs4d2}\;.
\ea 
Let us add a few comments, for the reader, about the methods used to derive that result. 
The term on the right hand side of Eq.~\nref{deltahiggs4d1} comes from $(++)$ mode contributions only, in the sense that it follows from the series
of Eq.~\nref{eq:che_fac} [second term of the whole expression],
\ba
 \sum_{q_1;d_1}(-\lm)\frac{\left(\al_{q_1d_1}\right)^2}{(M_{q_1}^2-\lm)(M_{d_1}^2-\lm)} \ ,  \label{eq:pureplusplusterm}
\ea
if one invokes the following identity,
\ba
\sum_{n=0}^\infty \frac{1}{n^2-x^2} = -\frac{1}{2x^2}\big[1+\left(\pi x\right)\cot\left(\pi x\right)\big]\; ,  \nonumber
\ea
where $x$ is some function of $R$ and $\lm$.
All the other terms of Eq.~\nref{eq:che_fac}, except the fifth one [last term of the second line] and the last one [two last lines], do not give contributions 
in the limit $\ep\to0$. The non-vanishing terms of Eq.~\nref{eq:che_fac} can be re-expressed as combinations of the (Hurwitz) Lerch 
transcendent~\footnote{Due to cancellations among different terms, the Eq.~\nref{deltahiggs4d1} does depend ultimately neither on $\g$ nor on $\log(-i\ep)$.},
\ba
 \Phi\left(e^{i\ep},1,x\right) & = & \sum_{n=0}^\infty \frac{e^{in\ep}}{n+x}\nn\\
 &=& -\gamma - \psi\left(x\right) - \log(-i\ep) + \mc O(\ep)\;, \nonumber
\ea
where $\g$ is the Euler-Mascheroni constant and $\psi\left(x\right) = \G^\prime(x)/\G(x)$
is the so-called digamma function (logarithmic derivative of the gamma function).

At this stage, we insist on the fact that in order to obtain Eq.~\nref{deltahiggs4d2} we have first written the CE of the mass matrix in Eq.~\nref{eq:che_fac}  
and calculated its KK summations up to $N\to\infty$, {\it before} imposing the limit $\ep\to0$ on the obtained CE -- as a last step. 
If, however, it is realized in the opposite order, i.e. first applying $\ep\to0$ on the mass matrix~\nref{eq:MassMat} [so that the matrix elements $\bt_{ij}\to 0$ 
since $q^n_R(\pi R) = d^n_L(\pi R)=0$], {\it before} writing the matrix CE and working out its infinite KK sums or in other words taking its limit for $N\to\infty$ 
[without $\bt_{ij}$ series anymore], one would obtain,
\ba
\tan^2\left(\pi R\sqrt{\vert m \vert^2}\right) &=&\left(\frac{vY_5}{\sqrt2}\right)^2 \label{deltahiggs4dpp}\;,
\ea
instead of Eq.~\nref{deltahiggs4d2}. Eq.~\nref{deltahiggs4dpp} originates solely from the series in Eq.~\nref{eq:pureplusplusterm}. 
As already observed in the 5D approach, one would obtain the same result as in Eq.~\nref{deltahiggs4dpp} by setting $Y^\prime_5=0$ in Eq.~\nref{deltahiggs4d2}~;
this is logical since the $\bt_{ij}$'s are proportional to $Y'_5$.

\subsection{Smoothing the Higgs peak}\label{sec:4d_sqdhiggs}

Alternatively, the Higgs Dirac peak at the boundary can be replaced by a normalized square function, of width $\ep R$, as in Eq.~\nref{eq:5deom_sq_a}-\nref{eq:5deom_sq_d}.
Then, using the profiles from Eq.~\nref{eq:4dprofiles}, we see that for $i,j>0$, 
\ba
\al_{ij} &=& \frac{vY_5}{\sqrt 2\ep R} \  \int_{(\pi-\ep)R}^{\pi R} dy \; q_L^i (y) \ d_R^j(y) \ = \ \frac{-vY_5}{\sqrt 2\ep R}\left(\frac{\sin[(i+j)(\pi-\ep)]}{i+j}+\frac{\sin[(i-j)(\pi-\ep)]}{i-j}\right) 
\ , \ \ \nonumber \\
\bt_{ji} &=&\frac{vY^\prime_5}{\sqrt 2\ep R} \  \int_{(\pi-\ep)R}^{\pi R} dy\; d_L^j (y) \ q_R^i(y) \ = \ \frac{-vY'_5}{\sqrt 2\ep R}\left(\frac{\sin[(i+j)(\pi-\ep)]}{i+j}-\frac{\sin[(i-j)(\pi-\ep)]}{i-j}\right) 
\ , \ \ \nonumber
\ea
which means that the functions $\al_{ij}$ and $\bt_{ji}$ are no longer factorizable in $i,j$ -- so that the simplification relation~(\ref{ASnature}) does not hold anymore. 
As a result, the CE of Eq.~(\ref{GenCE})-(\ref{eq:chpolP0})-(\ref{eq:chpolP1}), for the infinite $\left[M^\dag M\right]$ matrix, contains multiple infinite series which render 
difficult its simplification. Now in the absence of a compact form, like the one in Eq.~\nref{eq:che_fac}, it is tricky to solve the CE and work out the exact squared mass eigenvalues, 
$\lambda = \vert m \vert^2$.


\section{Interpretation of the analytical results}

\subsection{A new non-commutativity in the 4D approach}
\label{4Dcom}

After having presented our analytical results, we now discuss their impacts, one by one.
First, we have found that the 4D calculation gives rise to different fermion mass spectrum definitions
in the two orderings of the calculation~: first taking the limit
$\epsilon \to 0$ (Higgs localization) in the mass matrix~(\ref{eq:MassMat}) before writing the characteristic equation and apply the limit $N \to \infty$ 
(here $N$ refers generically to the various indices used in previous section for the KK summations), 
leads to the characteristic equation~(\ref{deltahiggs4dpp})~\footnote{In a preliminary work~\cite{Goertz:2008vr} on the RS framework, 
an approximated mass spectrum was obtained in this ordering ($\epsilon \to 0$, $N \to \infty$) through an expansion in powers of $v^2/M_{KK}^2$.}, 
while the inverse order of taking $N \to \infty$ in the characteristic equation and $\epsilon \to 0$ in a second step, results in Eq.~(\ref{deltahiggs4d2}) 
-- in the case of an Higgs profile regularized by a shifted Dirac peak where the characteristic equation can be derived analytically from the 4D point of view
(dealing with infinite mass matrices). In the former order, the fermion $(--)$ wave functions play absolutely no r\^ole in the calculation since the $\beta_{ij}$ off-diagonal terms 
of the matrix~(\ref{eq:MassMat}) vanish at the first step [$\epsilon \to 0$ limit]. In contrast, the infinite KK sum over these vanishing terms gives rise non-trivially 
to an additional contribution in Eq.~(\ref{deltahiggs4d2}) which is proportional to the $Y_5'$ coupling (entering the $\beta_{ij}$'s).
All this is summarized in the 4D line of Table~1.
\\
This non-commutativity will be confirmed in Section~\ref{Match} in the following sense: we will see that these two 4D calculation 
orderings correspond to two different 5D calculations.

$$ 
 \begin{array}{c|c|c}
\mbox{\bf Table 1}  \ \mbox{{\small\it (shifted Higgs)}} & \mbox{Regularization I} & \mbox{Regularization II}     \\
\hline
 & &   \\
& \boxed{\tan\left(\pi R \;m\right) = \frac{vY_5}{\sqrt2}} & \boxed{\tan\left(\pi R\; m\right) = \frac{\sqrt2(1+c)^2 vY_5}{2(1+c)^2+c v^2Y_5Y_5'}}  \\
\mbox{\underline{{\bf 5D} {\sc calculation}}}    & &   \\
 & \mbox{no $\delta$-terms for $(--)$-profiles} & \mbox{$\delta$-terms for $(--)$-profiles}   \\
 & \mbox{$(--)$ BC at $\pi R$, EOM with $\epsilon$} & \mbox{EOM with $\epsilon$, $(--)$ BC at $\pi R$} \\
 & &   \\
 \hline
  & &   \\
   & \boxed{\tan^2(\pi R\sqrt{\vert m\vert ^2}) = \left(\frac{vY_5}{\sqrt2}\right)^2} & \boxed{\tan^2(\pi R\sqrt{\vert m\vert ^2}) =\left(\frac{ vY_5/\sqrt2}{1+v^2Y_5Y_5'/8}\right)^2}     \\
 \mbox{\underline{{\bf 4D} {\sc calculation}}}   & &   \\
 & \mbox{no $(--)$-profile r\^ole} & \mbox{$(--)$-profile effect}  \\
 & \mbox{$\epsilon \to 0 \ , \ N \to \infty$} & \mbox{$N \to \infty \ , \ \epsilon \to 0$} 
  \end{array} 
$$ 
\begin{center}
 {\small \underline{Table 1~:} Quark mass spectrum for a shifted Higgs peak.}
\end{center}

In the context with bulk fermions coupled to a brane Higgs, the non-commutativity pointed out here -- the difference between the two orderings 
of the limits $\epsilon \to 0$ and $N \to \infty$ --
differs from the non-commutativity discussed in the literature~\cite{Malm:2013jia,Carena:2012fk} (in the RS framework)~: 
the latter one concerns the different results obtained from taking first 
$\epsilon \to 0$ and then $N_{KK} \to \infty$ as in Ref.~\cite{Casagrande:2010si}, or, the opposite order as in Ref.~\cite{Azatov:2010pf}. 
Here, $N_{KK}$ denotes the number of exchanged excited modes included at the level of the one-loop amplitude, when calculating the gluon-gluon 
fusion mechanism or the Higgs decay rate into two photons (the loop momentum integration is performed at the really first step). 
\\
While the 4D order $\epsilon \to 0$, $N_{KK} \to \infty$ matches the 5D calculation (avoiding the very notion of KK state) with a Higgs strictly stuck on the TeV-brane 
[where the $(--)$ KK modes vanish]~\cite{Malm:2013jia}, the opposite 4D order -- with the brane-limit taken only at last -- renders the Higgs sensitive to $(--)$ KK states 
and thus corresponds to the 5D approach with a narrow bulk-Higgs field localized towards the brane~\cite{Malm:2013jia} (unsuppressed `resonance contribution' 
from high-mass KK states which can resolve the Higgs wave function~\cite{Delaunay:2012cz}).
It was also found in Ref.~\cite{Malm:2013jia} that the limit $\epsilon \to 0$, for the Higgs profile regulator, can be taken either before 
or after performing loop integrations.
\\
Finally, let us comment that the question~\footnote{It was pointed out~\cite{Frank:2013un}, based on a 4D calculation of the gluon-gluon fusion amplitude in RS, 
that some specific higher derivative operators allow to take into account a UV-sensitivity.} about the non-commutativity of $\epsilon \to 0$ and $N_{KK} \to \infty$ 
has only a formal interest and was discussed for technical reasons since one has
to impose anyway a $\Lambda$ cut-off at the end of the day, due to the non-renormalizability of higher-dimensional theories (or their induced low gravity scale), 
so that $N_{KK}$ is bounded from above. By the way, it was found in Ref.~\cite{Malm:2013jia,Carena:2012fk} 
that once the loop calculation is performed in a realistic context with a consistent UV regulator such as dimensional regularization  
(or with a hard UV momentum cut-off on the 4D loop integral), the non-commutativity ambiguity disappears.
In contrast, the present non-commutativity of $\epsilon \to 0$ and $N \to \infty$ introduces indeed physical questions, because 
the $\Lambda$ cut-off must not be applied on $N$ (see Section~\ref{CutOff}).
These physical questions about the interpretation of the non-commutativity will be addressed in Section~\ref{scheme}.

\subsection{Matching the 4D and 5D approaches}
\label{Match}

In the 5D approach, there are also two possible ways for calculating the fermion mass spectrum, as described in Section~\ref{sec:5d} and  summarized
in the 5D line of Table~1.
\\
In one way, the BC at $\pi R$ is imposed for the $(--)$ profiles in a first stage so that the two terms in Eq.~(\ref{mmprofileD})-(\ref{mmprofileQ}) 
involving both the $\delta(y-\pi R)$ peak and a $(--)$ profile, $d_L(y)$ or $q_R(y)$, vanish (after integration).  
In a second stage, one solves the EOM system~(\ref{ppprofileD})-(\ref{ppprofileQ}) with a regularized Higgs peak, {\it e.g.} shifted by an amount $\epsilon R$. 
\\
The other way consists of first solving the system~(\ref{eq:5deom_delta_a})-(\ref{eq:5deom_delta_d}) with an $\epsilon R$-shifted Higgs, so that the terms 
in Eq.~(\ref{eq:5deom_delta_b})-(\ref{eq:5deom_delta_c}) involving both $\delta(y-(\pi-\epsilon) R)$ and a $(--)$ profile, $d_L$ or $q_R$, really contribute.
Then one imposes the BC at $\pi R$ for the $(--)$ profiles, which does not eliminate the above terms. The mass spectrum is dictated by those last conditions.
\\ 
Those two calculation orderings result in two different mass spectrum definition given by Eq.~(\ref{eq:5ddeltahiggspres}) 
and Eq.~(\ref{deltahiggs5dpp}), which are copied in the 5D line of Table~1~; the angle of the $\tan$ function is only defined modulo $n\pi$ 
which gives rise to the KK eigen-mass tower $m_n$ [$n\in \mathbb{N}$ as in Eq.~(\ref{first5Dfield})-(\ref{last5Dfield})]. The effect of the ElectroWeak symmetry 
breaking is thus a shift of $\arctan(vY_5/\sqrt2)/\pi R$ in the KK mass tower $n/R$, for the case of the left column in Table~1.

As expected~\footnote{In the 4D limit $N \to \infty$, the effect of the infinite KK tower is taken into account which is equivalent to consider rigorously the full 5D fields of 
Eq.~(\ref{Abasic5Dfield})-(\ref{Dbasic5Dfield}).}, 
there is a mass spectrum matching between the 4D and 5D calculations that Table~1 exhibits. Although expected, this matching was not trivial to demonstrate analytically, 
especially due to the complexity of dealing with the infinite 4D mass matrix~(\ref{eq:MassMat}). Furthermore, it turns out that there are in fact two distinct 4D/5D matchings, 
for the two calculation orders performed in 4D ({\it c.f.} Section~\ref{4Dcom}) and 5D (described in previous paragraph) that we thus commonly denote in the table 
as regularizations of type I and II -- see later discussion in the next Section~\ref{scheme}.
The 4D/5D matching in the regularization of type I is explicit~: the two equations obtained give rise to the same possible mass spectra.   
In the regularization of type II, the 4D/5D matching occurs exactly for $c=1$ as show the two mass equations~; it means that 
other 5D $c$-prescriptions [i.e. $c\neq 1$] do not represent physically distinct 
regularizations~\footnote{The precise notion of physically equivalent regularizations will be described at the beginning of Section~\ref{scheme}.} 
(as distinct 4D approaches matching $c\neq 1$ do not exist).

The first implication of those two 4D/5D matchings is the existence of two different 4D calculations (confirming subsection~\ref{4Dcom}) since there are two ways
of calculating the mass spectrum from the 5D point of view as well. These two ways of calculating (regularizations I and II) differ in their brane-Higgs sensitivity to 
the tower of bulk $(--)$ profiles~; this can be described remarkably in both the 4D and 5D approaches.
From the 5D point of view, in regularization II the terms in Eq.~(\ref{eq:5deom_delta_b})-(\ref{eq:5deom_delta_c}) coupling the VEV to $(--)$ profiles are not vanishing 
-- in contrast with case I -- as explained at the beginning of this subsection. Regarding the 4D treatment, in regularization II there is a non-vanishing contribution 
from the $\beta_{ij}$ terms [{\it c.f.} Eq.~(\ref{def:beta})] which represent overlaps between the Higgs and $(--)$ profiles, whereas their contribution is absent in case I as 
discussed at the beginning of Subsection~\ref{4Dcom}.

There is a second consequence~; the two 4D/5D matchings guarantee that the 5D mixed-formalism [{\it c.f.} Eq.~(\ref{first5Dfield})-(\ref{last5Dfield})], 
followed usually in literature, represents a correct procedure to take into account mixing effects between all KK levels which are otherwise {\it explicitly} 
included via the off-diagonal elements of the 4D mass matrix~(\ref{eq:MassMat}).

Finally, the 4D/5D matching confirms that there exist two approaches for deriving the same mass spectrum and that in the 4D approach there is no inconsistency
induced by the Higgs localization that should be regularized (as the so-called jump problem in the 5D approach). This can be interpreted by the fact that the exact 4D 
calculation proceeds by construction through a limit ($N\to \infty$) to obtain `softly' the fermion mass expressions in the wanted higher-dimensional scenario. 
This limit acts typically as the regularizing limit $\epsilon \to 0$ corresponding to a brane-Higgs, in the 5D framework.  
\\
The obtained 4D/5D matching also constitutes a confirmation of the theoretical validity of the field theory regularizations usually applied in the 5D calculation, 
and thus, leads to a global coherent picture. Now of course, to determine whether such a paradigm -- relying on mathematical re\-gularizations of an ill-defined peaked field -- 
corresponds really to the physical model, one would have to confront it with experimental results~\footnote{As the renormalizations of quantum corrections were 
confronted (with success) to collider data.}.

\subsection{On the two types of regularizations}
\label{scheme}

It is mentioned at the end of Appendix C.2 in Ref.~\cite{Csaki:2003sh} (where description is limited to the simpler case $Y_5=Y_5'$) 
that the regularizations, called I and II here, give at most two different interpretations of the $Y_5v(=yvR)$ parameter combination [proportional to $M_DL$ in notations
of Ref.~\cite{Csaki:2003sh}]. Let us discuss here this twofold feature more precisely. In fact, the two types of equations in Table~1 
(both similar in 4D and 5D for $c=1$) corresponding to the two regularizations 
constitute two different relations between the $Y^{(\prime)}_5$, $v$, $R$ parameters and the physical mass solutions represented by $m$. 
A physical mass $m$ having a unique value [the measured one], the difference between these two relations has to be either compensated 
by different values for $Y_5$, $v$, $R$ (which do not constitute observables) in cases I and II, or, cancelled by setting $Y'_5$ to zero 
(then $Y_5$, $v$, $R$ can be identical in cases I and II). 
There exist thus two numerically equivalent definitions of the mass value $m$ so that the regularizations I and II are physically equivalent
or even strictly identical [for vanishing $Y'_5$].
\\
Indeed, concretely, today there exist two different sets of $Y_5$, $v$, $R$ values (for $Y'_5\neq 0$) reproducing the measured values of the observed 
fermion masses through the two definitions, $f^{\rm I}_{n}$ and $f^{\rm II}_{n}$ (solutions from the two mass equations in Table~1), 
associated to the regularizations I and II~:
\begin{eqnarray}
\mbox{Regularization  I} \ \left \{ \begin{array}{l} 
m_{n}=f^{\rm I}_{n}(R,v,Y_5)  \\
\tilde m_{n}=f^{\rm I}_{n}(R,v,\tilde Y_5)  \end{array} \right .
\ \ \ 
\mbox{Regularization  II}   \ \left \{ \begin{array}{l} 
m_{n}=f^{\rm II}_{n}(R,v,Y_5,Y'_5)  \\
\tilde m_{n}=f^{\rm II}_{n}(R,v,\tilde Y_5,\tilde Y'_5)  \end{array} \right . \ 
\label{TwoRegulSystem}
\end{eqnarray}
In other words, the two systems in Eq.~(\ref{TwoRegulSystem}) have solutions in terms of $Y^{(\prime)}_5$, $v$, $R$ 
for the first mass eigenvalue [$m_{n=0}$] and this is true including quarks/leptons (same formalism as here introducing parameters 
$m_{\ell n}$, $Y_{\ell 5}$, $Y'_{\ell 5}$) of down or up ${\rm SU(2)_L}$-isospin 
(notations trivially extended to $\tilde m_{n}$, $\tilde Y_5$, $\tilde Y'_5$, $\tilde m_{\ell n}$, $\tilde Y_{\ell 5}$, $\tilde Y'_{\ell 5}$)  
from the three generations (notations to be completed with flavor indices). The fact that there exist solutions to the systems of type~(\ref{TwoRegulSystem}) 
is also due to the individual dependences of the masses on the Yukawa parameters~\footnote{Basically different 
masses depend on different $Y_5$-like parameters (i.e. $Y_5$, $\tilde Y_5$, $Y_{\ell 5}$,\dots).} 
and the higher number of $Y_5$-like parameters compared to the number of measured fermion masses. As for an overview of the other parameters, typically, 
the EW precision tests from the LEP collider would bound from above the $R$ radius (imposing large KK masses to avoid dangerous corrections to the SM 
predictions for EW observables) while in the gauge boson sector $m_Z$, $m_W$, $G_F$ would allow to determine 
the values of the bare parameters $v$, $g$, $g'$ (through loop calculations as described {\it e.g.} in Ref.~\cite{RSHiggs}), the recently measured Higgs mass 
fixing the quartic coupling $\lambda$~\cite{Djouadi:2005gi}~\footnote{Going from this toy model to RS~\cite{Randall:1999ee}, 
one should add the $AdS_5$ curvature parameter, $k$, but $kR\approx 11$ is
fixed by the gauge hierarchy solution. For the RS custodially protected version~\cite{ADMS} there can be an additional freedom from the $\tilde M$ parameter of explicit bulk 
custodial symmetry breaking, or even another one via the $g_{Z'}$ coupling~\cite{RSEWAFBb} in case of no Left-Right parity~\cite{O3}.
One should also add basically the 5D mass parameters, $c_{u,d,\nu,l}^{L/R \ i}$ ($i=1,2,3$), in the RS extensions addressing the flavor problem~\cite{RSmass}.}.                            
\\  
The physical equivalence of the regularizations I and II is based on generic arguments and thus also applies to amplitudes induced by Flavor Changing Neutral Current 
(FCNC) effects. This leads us to make new comments on Ref.~\cite{Azatov:2009na} which deals with FC Higgs couplings coming from misalignments between fermion masses 
and Yukawa couplings, in the RS framework with a brane Higgs. This misalignment is quantified by a non-universal shift estimated to be, using notations of Ref.~\cite{Azatov:2009na} 
except for down quark Yukawa parameters~:
\begin{eqnarray}
\begin{array}{l}
\mbox{Regularization  I}  \ \ \bigg \{  
\Delta^d = 0 + \Delta^d_2 = m_d \vert m_d\vert^2 R^{\prime 2} \bigg ( \frac{F(c_q)}{f(c_q)^2}+\frac{F(-c_d)}{f(-c_d)^2} \bigg )  
\\ 
\mbox{Regularization  II} \ \ \bigg \{  
\Delta^d = \Delta^d_1 + \Delta^d_2 = m_d \vert m_d\vert^2 R^{\prime 2} \bigg ( \frac{2}{3} \frac{Y'_5}{Y_5}  \frac{1}{f(c_q)^2f(-c_d)^2}  +  \frac{F(c_q)}{f(c_q)^2}+\frac{F(-c_d)}{f(-c_d)^2}  \bigg )  
\end{array}
\label{FCshift}
\end{eqnarray}
where $F(c_q) = (2c_q-1)/(2c_q+1)$ and $\Delta^d_1=0$ in case I due to vanishing contributions from $Y'_5$ terms. 
Note that in these equations the physical condition to reproduce the (approximated) $m_d$ mass has been used to fix the $v$ parameter.
Eq.~(\ref{FCshift}) shows that there exist two sets of parameters~\footnote{For instance with $f(c_q)\ll 1$, $f(-c_d)\ll 1$ and $Y'_5\ll Y_5$.} 
giving rise to the same value of $\Delta^d$ within the regularizations I (without terms proportional to $Y'_5$, as included in Ref.~\cite{Azatov:2009na}) 
and II (with such terms) so that these regularizations can be physically equivalent. 
There even exist such parameters ({\it e.g.} $f(c_q) \sim 1$, $f(-c_d)\ll 1$) for $Y'_5$ and $Y_5$ of the same order of magnitude as might be wanted to not introduce 
new energy scales~\cite{Azatov:2009na}. Notice that with more constraints on parameters from new experimental data and under the strong physical assumption $Y'_5 \simeq Y_5$, 
it could happen that the two sets of input parameters in regularizations I and II cannot reproduce the same value of $\Delta^d$~: then precise FCNC data should be used to 
select the correct theoretical regularization by pinning down the real and unique $\Delta^d$ value. This experimental test is similar to the one discussed right below.

In the future, the upgraded 13~TeV LHC and other colliders will certainly provide more data. One can expect more precise measurements of the Yukawa  
and $hVV$ [$V=Z,W$] couplings (being functions of $g$, $g'$, $v$~\cite{Djouadi:2005gi} and $R$ due to KK gauge boson mixings) or even the detection of 
Higgs pair production that would give information on the $hVV$, $hhVV$, $hhh$ couplings (in turn on combinations of $\lambda$, $g$, $g'$, $v$, $R$). 
The systems of Eq.~(\ref{TwoRegulSystem}) would thus have to be extended to include in particular the physical $Y_{nm}$, $\tilde Y_{nm}$ 
Yukawa couplings which depend on the same parameters $Y^{(\prime)}_5$, $\tilde Y^{(\prime)}_5$, $v$, $R$~:
\begin{eqnarray}
\mbox{Regularization  I}  \ \left \{ \begin{array}{l} 
m_{n}=f^{\rm I}_{n}(R,v,Y_5)  \\
\tilde m_{n}=f^{\rm I}_{n}(R,v,\tilde Y_5)  \\
Y_{nm}=g^{\rm I}_{nm}(R,v,Y_5)  \\
\tilde Y_{nm}=g^{\rm I}_{nm}(R,v,\tilde Y_5)  \end{array} \right .
\ \ \ 
\mbox{Regularization  II} \ \left \{ \begin{array}{l} 
m_{n}=f^{\rm II}_{n}(R,v,Y_5,Y'_5)  \\
\tilde m_{n}=f^{\rm II}_{n}(R,v,\tilde Y_5,\tilde Y'_5)  \\
Y_{nm}=g^{\rm II}_{nm}(R,v,Y_5,Y'_5) \\
\tilde Y_{nm}=g^{\rm II}_{nm}(R,v,\tilde Y_5,\tilde Y'_5)  \end{array} \right . \ 
\label{TwoRegulSystemFut}
\end{eqnarray}
Those couplings are involved in the action terms $Y_{nm}h(x)\bar Q^n_L(x)D^m_R(x)$ and $\tilde Y_{nm} h(x)\bar {\tilde Q}^n_L(x)U^m_R(x)$ expressed with 4D fields 
representing mass eigenstates~\footnote{In the 4D approach, this notation is coherent with previous notations if the $D^m_R(x)$ fields result from a mixing with the 
$Q^m_R(x)$ fields, and the $Q^n_L(x)$ include mixings with $D^n_L(x)$.}. KK mode discoveries would also add new entries (like $m_n$ with $n>0$) for the systems 
in Eq.~(\ref{TwoRegulSystemFut}). 
\\
With such new data coming it could happen at some point that there exist no more set of parameters satisfying one of the two types of system in Eq.~(\ref{TwoRegulSystemFut}) 
[more physical constraints without new degrees of freedom]. This would mean that the associated regularization is ruled out by experimental data. 
This uniquely ruled out regularization could only correspond to the system with less parameters~: regularization I (no $Y'_5$, $\tilde Y'_5$ parameters), 
since regularization II for $Y'_5,\tilde Y'_5\to 0$ gives back regularization I so that excluding regularization II would also exclude regularization I.
In a situation of this kind where the regularization I only is experimentally ruled out, the regularizations I and II would obviously not be physically equivalent.
\\
Let us simply remark here that it is not trivial to conclude {\it intuitively} on the physical equivalence of the two regularizations. Indeed in regularization II, 
from the 4D point of view, first taking $N \to \infty$ leads to have in a first step a full 5D theory with complete (i.e. infinite) 5D field KK decompositions. 
Then imposing the $\epsilon \to 0$ limit, in this non-truncated 5D framework, represents effectively a localization of the Higgs scalar on the brane. 
In contrast, for the regularization I, the physical sense of taking $\epsilon \to 0$ before having completed the 5D theory 
(i.e. having taken $N \to \infty$) is not clear anymore~: it is not obvious that it corresponds to the geometric brane-localization along the extra-dimension 
as it is realized within an hybrid 5D scenario. In other words, this regularization may or may not be equivalent to regularization II.  
Therefore the experimental tests described above are really necessary to determine whether those two regularizations are physically equivalent or not.

The above considerations on the degrees of freedom added by the $Y'_5$, $\tilde Y'_5$ parameters are expected to be similar with a warped extra-dimension.
Therefore, one can invoke the previous discussion to make the following comments on the past and future literature about the RS scenario (or generally on 
higher-dimensional theories with a brane-localized Higgs scalar and bulk matter)~\footnote{For constructions of RS scenarios with a brane-Higgs as a limit  
case of bulk-Higgs models, we refer to Ref.~\cite{Azatov:2010pf,Azatov:2009na}.}.
\\
As discussed at the beginning of this subsection, the regularizations I and II reproduce the present collider data and are thus physically equivalent.
Hence, the constructions of RS realizations reproducing the fermion masses and mixings performed through the regularization I, as 
for instance in Ref.~\cite{RSEWAFBb,RSmass}, would have been possible as well using regularization II. 
\\
Concerning future data, one cannot be sure to predict theoretically all the possible physical values within regularization I [some can be inaccessible as discussed 
below Eq.~(\ref{TwoRegulSystemFut})] whereas regularization II is clearly exhaustive in its predictions (it includes the parameter space of regularization I which 
is recovered for $Y'_5=\tilde Y'_5= 0$). 
This is the reason why the RS predictions on KK quark masses, FCNC rates or Higgs productions/decays (involving KK fermion mixings) made {\it e.g.} in 
Ref.~\cite{RSfusion,RSHiggs,AgS,ABP,APS} (4D calculation) \cite{Casagrande:2008hr} (5D calculation)~\footnote{Let 
us also mention Ref.~\cite{Goertz:2008vr} in regularization I, which presents 4D/5D matching considerations via a numerical approach and for a truncated KK fermion tower.} 
may not be complete in contrast with those of Ref.~\cite{RSggH5D,Malm:2013jia,Carena:2012fk,Casagrande:2010si,Azatov:2009na} (5D calculation).     
\\
Finally, our recommendations to treat the future experimental data within the RS model are as follows. One should perform the regularizations I and II to determine whether in both 
cases there exist parameters reproducing the whole set of observables [as in Eq.~(\ref{TwoRegulSystemFut})]. If the regularization I cannot reproduce data then it is excluded, 
otherwise the two regularizations are physically equivalent~\footnote{The last possible situation with both regularizations unable to reproduce data would mean that either another 
kind of regularization is necessary or the RS model itself (in its minimal version with an Higgs boson strictly localized on the brane) is ruled out.}. This procedure is important to 
safely conclude on the validity of these Higgs regularizations and to avoid misleading interpretations. 
From a practical point of view, the question of the phy\-sical equivalence of these regularizations is also important. Indeed, a systematic calculation of the fermion masses or Yukawa 
couplings is easier through regularization I than II, both in the 4D [less infinite sums to address cause some mass  
matrix elements vanish] and 5D [less $\delta(y-\pi R)$ terms in EOM's] approaches. Therefore, one could benefit from a regularization equivalence by choosing 
to use the simpler regularization I.

\subsection{The correct cut-off procedure}
\label{CutOff}

Generally speaking, the extra-dimensional backgrounds lead to non-renormalizable theories which are valid only up to a certain energy scale where starts the 
non-perturbative regime. For instance, in the RS model with bulk matter this scale is driven by the perturbativity of the top Yukawa coupling
and is around $2$-$3 M_{KK}$ ($M_{KK}\equiv$~first KK photon mass) [see {\it e.g.} Ref.~\cite{RSEWAFBb}] so that a $\Lambda$ cut-off satisfying, 
$\Lambda \lesssim 2$-$3 M_{KK}$, should be applied. $\Lambda$ indicates the typical energy scale of the UV completion of the theory.

Based on the previous results and discussions, 
we are going to clarify here the correct and generic way to apply the $\Lambda$ cut-off on scenarios with a Higgs scalar stuck at a brane.  
Without loss of generality, one should follow this two-step procedure, 
\\ \\
{\it {\bf (1) :} calculate the bulk fermion mass spectrum and Yukawa couplings including infinite KK tower contributions, as done automatically when manipulating 
5D fields or considering infinite mass matrices [with $N \to \infty$ after/before $\epsilon \to 0$ accordingly to regularization I/II] in the 4D approach,
\\ 
{\bf (2) :} consider only the obtained mass eigenstates of the towers [masses and couplings derived at step} (1){\it ] which are lighter 
than the $\Lambda$ cut-off, in the computation of physical observables and tree/loop-level amplitudes -- 
with notations of Section~\ref{4Dcom}, it means that $N_{KK}$ must be finite~\footnote{Even if a cut-off should be applied on physical observables,  
it may be instructive to take the limit $N_{KK}\to \infty$ for technical purposes in formal discussions on the calculation 
itself~\cite{RSggH5D,Malm:2013jia,Carena:2012fk,Casagrande:2010si,Azatov:2010pf}.}.}
\\ \\
The reason for this rigorous order is that one should {\it first} build formally a consistent and pure 5D theory ($N \to \infty$) with full KK fermion mixings, 
{\it before} truncating this theory at the frontier of its validity domain indicated by $\Lambda$ to get the physical effective low-energy model. 
\\
Notice that adopting the inverse order, i.e. (2)$\to$(1), within regularization II, that is first applying the $\Lambda$ cut-off and secondly  
calculate the fermion mass eigenvalues with a finite mass matrix (as the cut-off would prevent from taking $N \to \infty$) -- ending with $\epsilon \to 0$ -- would 
lead to incomplete eigen-mass expressions (even for the lightest modes) without the $Y'_5$ term [{\it c.f.} Table~1]. Indeed, the non-vanishing 
contributions from the mass matrix elements involving $Y'_5$ originate non-trivially 
from the fact that the limit $N \to \infty$ has been taken [see beginning of Section~\ref{4Dcom}].

This cut-off procedure is analog in supersymmetric RS extensions~\cite{RSUSY} where, at the first step, the 4D effective Lagrangian must be written including infinite KK tower effects~: 
one can then regularize tree-level $\delta(0)$-inconsistencies, arising in the bulk sfermion couplings to two brane-Higgs bosons (from Yukawa and D-terms)~\footnote{And in self couplings
of the Higgs bosons as well.}, through cancellations with contributions from exchanges of infinite KK towers -- treated via the completeness relation. In a second step, one can apply the 
$\Lambda$ cut-off on tower eigenstates entering the computation {\it e.g.} of quantum corrections to the Higgs mass, based on the obtained couplings~\cite{RSUSY}. 
This procedure, which has been shown to be the correct one in supersymmetric RS frameworks~\cite{RSUSY}, confirms that one should first elaborate a consistent and thus complete
5D theory (with infinite KK towers) {\it before} truncating it at the physical cut-off for calculating amplitudes -- as justified in previous paragraph.

\subsection{Discussion for the square Higgs profile}
\label{AddRem}

Let us finally discuss the regularization introduced in Section~\ref{5Dbox}, which consists in smoothing the Higgs delta peak by a square function. In that case, 
depending on whether the $(--)$ BC at $\pi R$ is applied before or after solving the EOM system~(\ref{eq:5deom_sq_a})-(\ref{eq:5deom_sq_d}) with a square Higgs 
profile, the mass spectrum is given by Eq.~(\ref{deltahiggs5dpp}) [regularization I] or Eq.~(\ref{SpecBox5DII}) [regularization II]. 
In the regularization I, there are no $\Tha$-terms for $(--)$-profiles in Eq.~(\ref{eq:5deom_sq_b})-(\ref{eq:5deom_sq_c}).
All this is summarized in Table~2 below, similarly to 5D part of previous Table~1 for the shifted Higgs regularization -- except that here $Y_5=Y'_5$ is assumed 
(case II) for simplicity in the calculation.
\\
The Higgs regularizations via a square profile and a shifted delta peak are physically equivalent~\cite{Azatov:2009na,Csaki:2003sh} for the same reasons 
as those presented in details at the beginning of Section~\ref{scheme} where regularizations I and II were compared. 
Note that in case of regularization I, these two profile re\-gularizations are even formally equivalent as show the identical mass spectrum exhibited in Tables~1 and 2.
Hence, the above discussion on the equivalence of regularizations I and II (Section~\ref{scheme}) hold also for the square Higgs profile. In particular, the considerations on the
counting of degrees of freedom are the same~: there are once more additional parameters ($Y'_5$) in regularization II [even if those do not appear explicitly in Table~2 due to the 
$Y_5=Y'_5$ hypothesis]. Finally, the discussion on the cut-off in Section~\ref{CutOff} remains also valid with a square Higgs profile.

$$ 
 \begin{array}{c|c|c}
\mbox{\bf Table 2}  \ \mbox{{\small\it (square Higgs)}} & \mbox{Regularization I} & \mbox{Regularization II \ \ \small{[$Y_5=Y'_5$]}}     \\
\hline
 & &   \\
& \boxed{\tan\left(\pi R \;m\right) = \frac{vY_5}{\sqrt2}} & \boxed{\tan\left(\pi R\;m\right)=\tanh\left(\sqrt{\frac{v^2Y^2_5}{2}}\right)}  \\
\mbox{\underline{{\bf 5D} {\sc calculation}}}    & &   \\
 & \mbox{no $\Tha$-terms for $(--)$-profiles} & \mbox{$\Tha$-terms for $(--)$-profiles}   \\
 & \mbox{$(--)$ BC at $\pi R$, EOM with $\epsilon$} & \mbox{EOM with $\epsilon$, $(--)$ BC at $\pi R$} \\
 & &   
  \end{array} 
$$
\begin{center}
 {\small \underline{Table 2~:} Quark mass spectrum for a square Higgs profile.}
\end{center}


\section{Summary and conclusions}

In the framework of a simple higher-dimensional model with bulk matter and a brane-localized Higgs boson~\footnote{Our results are expected to be essentially similar 
(up to warp factors of course) in realistic warped extra-dimension scenarios.}, we have first pointed out a certain non-commutativity 
in the order of the 4D calculation for the fermion mass spectrum~: applying first the limit $\epsilon \to 0$ and then $N \to \infty$ (so-called regularization I) leads to a different
analytical expression from the inverse ordering (regularization II). The interpretation of this difference raises obviously a new physical question~: which order is correct~? 
Note that this non-commutativity differs from the one recently addressed in literature~\cite{Malm:2013jia,Carena:2012fk} (in the RS model) which concerns the 
limits $\epsilon \to 0$ and $N_{KK} \to \infty$, $N_{KK}$ denoting the number of KK modes exchanged in Higgs production/decay amplitudes.

Then the exact matching between the 4D and 5D calculations of the mass spectrum, which is expected, has been established analytically 
-- for the first time and in both regularizations (I/II). This matching allows a deeper understanding of the regularizations of brane-Higgs models~;
in particular, it turns out that the regularizations I and II differ in their brane-Higgs sensitivity to the tower of bulk $(--)$ profiles for the fermions. 
Besides, the obtained 4D/5D matching represents a confirmation that the usually applied 5D mixed-formalism [i.e. the mixed KK decomposition of Eq.~(\ref{first5Dfield})-(\ref{last5Dfield})] 
is a correct way of including the whole KK mixing effect.

We have further worked out the interpretation of the existence of two types of Higgs peak regularization, which answers the question previously raised about the new 
non-commutativity. The conclusion is that with the present experimental setup, the regularizations I and II are physically equivalent. Nevertheless, with future constraints from 
high-energy collider results, it could happen that only the regularization I is ruled out -- as the regularization II involves more free parameters (like $Y'_5$, $\tilde Y'_5$). 
\\
Based on these considerations, our recommendations are to use regularization II for making complete predictions on measurable observables,  
and, to compare the two regularizations when confronting the extra-dimensional model to new data in order to conclude about their physical equivalence. Such a conclusion 
will avoid dangerous misleading interpretations on regularizations and might prove to be useful, from a technical point of view, in the choice of the calculation method.

Our analysis has lead us to clarify the cut-off procedure in models with a brane-Higgs~: one must {\it first} build a consistent 5D theory -- i.e. calculate eigen-masses and Yukawa
couplings accordingly to regularization I or II -- with full KK fermion effects ($N \to \infty$), {\it before} restricting this theory (finite $N_{KK}$) to its validity domain delimited by the 
$\Lambda$ UV cut-off for computing physical amplitudes. This is analog to the cut-off process in supersymmetric extensions of the RS model~\cite{RSUSY}.

Finally, we mention that even if the Higgs peak regularization used throughout the paper was shifting the delta peak, regularizing the Higgs profile by a smooth square function
is physically equivalent and has been performed as well in this paper. In particular, this square profile treatment has allowed to confirm our statements on the comparison 
between the regularizations I and II.


\section*{Acknowledgements}

The authors would like to thank Andrei Angelescu, Michel Dubois-Violette, 
Belen Gavela, Jose Santiago and Manuel Toharia for motivating and interesting discussions.
The works of R.~B. and S.~M. are supported respectively by the P2IO Labex and CNRS. 
G.~M. acknowledges support from the ERC Advanced Grant Higgs@LHC and 
the {\it Institut Universitaire de France (IUF)}.


\appendix

\renewcommand\thesection{Appendix}      
\renewcommand\thesubsection{\Alph{section}.\arabic{subsection}}
\renewcommand\thesubsubsection{\Alph{section}.\arabic{subsection}.\arabic{subsubsection}}

\renewcommand{\theequation}{\Alph{section}.\arabic{equation}}     
\renewcommand{\thetable}{\Alph{section}.\arabic{table}}           
\renewcommand{\thefigure}{\Alph{section}.\arabic{figure}}        

\setcounter{section}{0}
\setcounter{figure}{0}
\setcounter{table}{0}
\setcounter{equation}{0}

\section{The generic characteristic equation}\label{sec:CE}

From the infinite quark mass matrix, $\left[M\right]$, defined in Eq.~\nref{eq:MassMat}, we first get the symmetric matrix, $\left[M^\dag M\right]$, which can be written 
without lost of generality as,
{\small  \ba
\left( \begin{array}{cccccc}  
M_{q_0}^2 +\sum_{n} \bt_{n0}^2 &\al_{00}M_{q0}+\bt_{00}M_{d0} & \sum_{n}\bt_{n0}\bt_{n1} & \al_{01}M_{q0}+\bt_{10}M_{d1}& \sum_{n}\bt_{n0}\bt_{n2}  & \cdots \\
\vdots & M_{d_0}^2 +\sum_{n} \al_{n0}^2& \al_{10}M_{q1}+\bt_{01}M_{d0} & \sum_{n}\al_{n0}\al_{n1}&\al_{20}M_{q2}+\bt_{02}M_{d0}&\cdots\\
\vdots & \vdots &M_{q_1}^2 +\sum_{n} \bt_{n1}^2  & \al_{11}M_{q1}+\bt_{11}M_{d1} & \sum_{n}\bt_{n1}\bt_{n2}  &\cdots\\
\vdots & \vdots & \vdots & M_{d_1}^2 +\sum_{n} \al_{n1}^2 & \al_{21}M_{q2}+\bt_{12}M_{d1}&\cdots\\
\vdots & \vdots & \vdots & \vdots & M_{q_2}^2 +\sum_{n} \bt_{n2}^2  & \cdots\\
\vdots & \vdots & \vdots & \vdots & \vdots & \ddots
\end{array} \right)
\nonumber
\ea }
where the discrete sums are taken from $n=0$ to infinity.

For the characteristic equation of this infinite matrix, $\left[M^\dag M\right]$, we find the following analytical expression,
\ba
1+ \mc P_0 + \mc P_1=0  \label{GenCE}
\ea
where $\mc P_0$ and $\mc P_1$ are both series of infinite number of terms. In the limit where all the KK mass $(M's)$ terms  
go to zero, $\mc P_1$ vanishes but $\mc P_0$ does not. In order to describe the parts, $\mc P_0$ and $\mc P_1$, we have to first define the following structures
depending on the $\al_{qd}$ [defined in Eq.~(\ref{def:alpha})],
\ba
A_n\left(q_1,\cdots, q_n; d_1,\cdots, d_n\right) \ \equiv \sum_{r_1,\cdots,r_n (\in \{d_1,\cdots, d_n\})}  \varepsilon^{i(r_1)...i(r_n)}
\ \al_{q_1r_1}\cdots\al_{q_nr_n}   \  ,
\ea
such that $\varepsilon^{abc...}$ is the anti-symmetric tensor, with for instance the index $i(r_3) = i(d_2) \hat = 2$, and the three first structures can be written explicitly as,
\bas
A_1\left(q_1; d_1\right) &=& \al_{q_1d_1} \  ,\\
A_2\left(q_1,q_2; d_1,d_2\right) &=& \al_{q_1d_1}\al_{q_2d_2} - \al_{q_1d_2}\al_{q_2d_1} \ ,\\
A_3\left(q_1,q_2,q_3; d_1,d_2,d_3\right) &=& 
\al_{q_1d_1}\al_{q_2d_2}\al_{q_3d_3}  + \al_{q_1d_2}\al_{q_2d_3}\al_{q_3d_1}
+ \al_{q_1d_3}\al_{q_2d_1}\al_{q_3d_2}\\
&& - \al_{q_1d_1}\al_{q_2d_3}\al_{q_3d_2}
-\al_{q_1d_3}\al_{q_2d_2}\al_{q_3d_1} -\al_{q_1d_2}\al_{q_2d_1}\al_{q_3d_3} \ .
\eas
Note that because of the anti-symmetric nature of these structures, in a factorizable case where $\al_{qd}=f_q \times f_d$, one has simply,
\ba
A_{n \geq 2} = 0. \label{ASnature}
\ea
Analog structures can be introduced for the $\beta_{dq}$ [defined in Eq.~(\ref{def:beta})]~:
\ba
B_n\left(d_1,\cdots, d_n; q_1,\cdots, q_n\right) \ \equiv \sum_{r_1,\cdots,r_n ( \in \{d_1,\cdots, d_n\})} \varepsilon^{i(r_1)...i(r_n)}
\ \bt_{r_1q_1}\cdots\bt_{r_nq_n}  \ . 
\ea
One has similarly, $B_{n \geq 2} = 0$, for factorizable cases with, $\bt_{dq}=f_d \times f_q$.
Now, with the conditions
\ba
q_i \neq q_j \neq Q_k &;&  d_i \neq d_j \neq D_k
\ea
where $i,j,k = 0,1,2, \cdots$, the first few terms (sufficient to deduce the rest of the infinite series) in the $\mc P_0$ and $\mc P_1$ parts can be expressed as,
\ba
\mc P_0 &=& \sum_{q_1;d_1}(-\lm)\left(\frac{\left(A_1(q_1;d_1)\right)^2}{(M_{q_1}^2-\lm)(M_{d_1}^2-\lm)} +(\al \leftrightarrow \bt)\right)\nn\\
&+&\sum_{q_1<q_3;d_1<d_3}(-\lm)^2\left(\frac{\left(A_2(q_1,q_3;d_1,d_3)\right)^2}
{(M_{q_1}^2-\lm)(M_{d_1}^2-\lm)(M_{q_3}^2-\lm)(M_{d_3}^2-\lm)} +(\al \leftrightarrow \bt)\right)\nn\\
&+&\sum_{q_1,q_2;d_1,d_2}(-\lm)^2 \ 
\frac{\left(A_1(q_1;d_1)\right)^2}{(M_{q_1}^2-\lm)(M_{d_1}^2-\lm)}\times \frac{\left(B_1(d_2;q_2)\right)^2}{(M_{q_2}^2-\lm)(M_{d_2}^2-\lm)}\nn\\
&&\times\left(1-\dl_{q_1q_2}\frac{M_{q_2}^2}{\lm}\right)
\left(1-\dl_{d_1d_2}\frac{M_{d_2}^2}{\lm}\right)\nn\\
&+&\sum_{q_1<q_3,q_2;d_1<d_3,d_2}(-\lm)^3\left\{\frac{\left(A_2(q_1,q_3;d_1,d_3)\right)^2}
{\prod_{i=1,3}(M_{q_i}^2-\lm)(M_{d_i}^2-\lm)}\times\frac{\left(B_1(d_2;q_2)\right)^2}{(M_{q_2}^2-\lm)(M_{d_2}^2-\lm)}\right.\nn\\
&&\times\left(1-\dl_{q_1q_2}\frac{M_{q_2}^2}{\lm}\right)
\left. \left(1-\dl_{d_1d_2}\frac{M_{d_2}^2}{\lm}\right)\left(1-\dl_{q_3q_2}\frac{M_{q_2}^2}{\lm}\right)
\left(1-\dl_{d_3d_2}\frac{M_{d_2}^2}{\lm}\right)+(\al \leftrightarrow \bt)\right\}\nn
\ea
\ba
&+&\sum_{q_1<q_3,q2<q_4;d_1<d_3,d_2<d_4}(-\lm)^4 \ 
\frac{\left(A_2(q_1,q_3;d_1,d_3)\right)^2}
{\prod_{i=1,3}(M_{q_i}^2-\lm)(M_{d_i}^2-\lm)}\times\frac{\left(B_2(d_2,d_4;q_2,q_4)\right)^2}{\prod_{j=2,4}(M_{q_j}^2-\lm)(M_{d_j}^2-\lm)}\nn\\
&&\times
\left(1-\dl_{q_1q_2}\frac{M_{q_2}^2}{\lm}\right)
\left(1-\dl_{q_3q_2}\frac{M_{q_2}^2}{\lm}\right)
\left(1-\dl_{q_1q_4}\frac{M_{q_4}^2}{\lm}\right)
\left(1-\dl_{q_3q_4}\frac{M_{q_4}^2}{\lm}\right)\nn\\
&&\times
\left(1-\dl_{d_1d_2}\frac{M_{d_2}^2}{\lm}\right)
\left(1-\dl_{d_3d_2}\frac{M_{d_2}^2}{\lm}\right)
\left(1-\dl_{d_1d_4}\frac{M_{d_4}^2}{\lm}\right)
\left(1-\dl_{d_3d_4}\frac{M_{d_4}^2}{\lm}\right)\nn\\
&+& \mc O(\al^6,\bt^6)\label{eq:chpolP0} \ , \ea
and, 
\ba
\mc P_1&=&\sum_{Q_1;D_1} \frac{-2 M_{Q_1}M_{D_1}}{(M_{Q_1}^2-\lm)(M_{D_1}^2-\lm)}\times\Bigg[ A_1(Q_1;D_1)\times B_1(D_1;Q_1)\nn\\
&&+\sum_{q_1;d_1} (-\lm)\left\{\frac{A_2(Q_1,q_1;D_1,d_1)A_1(q_1;d_1)
}{(M_{q_1}^2-\lm)(M_{d_1}^2-\lm)}\times B_1(D_1;Q_1) + (\al \leftrightarrow \bt)\right\}\nn\\
&&+\sum_{q_1q_2;d_1,d_2} (-\lm)^2\times \frac{A_2(Q_1,q_1;D_1,d_1)A_1(q_1;d_1)}{(M_{q_1}^2-\lm)(M_{d_1}^2-\lm)}
\times\frac{B_2(D_1,d_2;Q_1,q_2)B_1(d_2;q_2)}{(M_{q_2}^2-\lm)(M_{d_2}^2-\lm)} \nn\\
&&\times\left(1-\dl_{q_1q_2}\frac{M_{q_2}^2}{\lm}\right)
\left(1-\dl_{d_1d_2}\frac{M_{d_2}^2}{\lm}\right)+ \mc O\left(\al^5, \bt^5\right)\Bigg]
\nn\\~\nn\\
&+&\Bigg\{\sum_{Q_1<Q_2}  \frac{2 M_{Q_1}M_{Q_2}}{(M_{Q_1}^2-\lm)(M_{Q_2}^2-\lm)}\times\Bigg[ \sum_{d_1,d_2} (-\lm)\times \frac{A_1(Q_1;d_1)A_1(Q_2;d_1)}{(M_{d_1}^2-\lm)}\nn\\
&&\times\frac{ B_1(d_2;Q_1)B_1(d_2;Q_2)}{(M_{d_2}^2-\lm)}\left(1-\dl_{d_1d_2}\frac{M_{d_2}^2}{\lm}\right)\nn\\
&&+\sum_{q_1;d_1,d_2,d_3}(-\lm)^2 \left\{\frac{A_2(Q_1,q_1;d_1,d_3)
A_2(Q_2,q_1;d_1,d_3)}{(M_{q_1}^2-\lm)(M_{d_1}^2-\lm)}\times\frac{B_1(d_2;Q_1)B_1(d_2;Q_2)}{(M_{d_3}^2-\lm)}\right.\nn\\
&&\times\left.
\left(1-\dl_{d_1d_2}\frac{M_{d_2}^2}{\lm}\right)\left(1-\dl_{d_3d_2}\frac{M_{d_2}^2}{\lm}\right)+ (\al\leftrightarrow \bt)\right\}\nn\\
&&+\sum_{q_1,q_2;d_1,d_2,d_3,d_4}(-\lm)^3\times\frac{A_2(Q_1,q_1;d_1,d_3)
A_2(Q_2,q_1;d_1,d_3)}{(M_{q_1}^2-\lm)(M_{d_1}^2-\lm)(M_{d_3}^2-\lm)}\nn\\
&& \times\frac{B_2(d_2,d_4;Q_1,q_2)B_2(d_2,d_4;Q_2,q_2)}{(M_{q_2}^2-\lm)(M_{d_2}^2-\lm)(M_{d_4}^2-\lm)}\left(1-\dl_{q_1q_2}\frac{M_{q_2}^2}{\lm}\right)
\left(1-\dl_{d_1d_2}\frac{M_{d_2}^2}{\lm}\right)\nn\\
&&\times\left(1-\dl_{d_1d_4}\frac{M_{d_4}^2}{\lm}\right)
\left(1-\dl_{d_3d_2}\frac{M_{d_2}^2}{\lm}\right)\left(1-\dl_{d_3d_4}\frac{M_{d_4}^2}{\lm}\right)+ \mc O\left(\al^6, \bt^6\right)\Bigg] \nn\\
\nn\\
&&+\left[(Q_1,Q_2) \rightarrow (D_1,D_2)\right]\Bigg\}
\nn
\ea
\ba
&+&\sum_{Q_1<Q_2;D_1<D_2} 2\left(\prod_{i=1,2}\frac{ M_{Q_i}M_{D_i}}{(M_{Q_i}^2-\lm)(M_{D_i}^2-\lm)}\right)\nn\\
&&\times\Bigg[
A_2(Q_1,Q_2;D_1,D_2)\times B_2(D_1,D_2;Q_1,Q_2) \nn\\
&& + \ A_1(Q_1;D_1)A_1(Q_2;D_2)\times B_1(D_1;Q_1)B_1(D_2;Q_2) \nn\\
&&+  \ A_1(Q_1;D_2)A_1(Q_2;D_1)\times B_1(D_2;Q_1)B_1(D_1;Q_2)
+ \mc O\left(\al^4, \bt^4\right)\Bigg] \nn\\
&+&\Bigg\{\sum_{Q_1<Q_2<Q_3;D_1} (-2)\left(\prod_{i=1,3}\frac{ M_{Q_i}}{(M_{Q_i}^2-\lm)}\right)\frac{M_{D_1}}{(M_{D_1}^2-\lm)}\nn\\
&&\times\Bigg[\sum_{d_1,d_2}(-\lm)\Bigg(
\frac{A_2(Q_1,Q_2;D_1,d_1 )A_1(Q_3;d_1)}{(M_{d_1}^2-\lm)}\times
\frac{B_2(D_1,d_2;Q_1,Q_2)B_1(d_2;Q_3)}{(M_{d_2}^2-\lm)}\nn\\
&&+\frac{A_2(Q_1,Q_3;D_1,d_1 )A_1(Q_2;d_1)}{(M_{d_1}^2-\lm)}\times
\frac{B_2(D_1,d_2;Q_1,Q_3)B_1(d_2;Q_3)}{(M_{d_2}^2-\lm)}\nn\\
&&+\frac{A_2(Q_2,Q_3;D_1,d_1 )A_1(Q_1;d_1)}{(M_{d_1}^2-\lm)}\times
\frac{B_2(D_1,d_2;Q_2,Q_3)B_1(d_2;Q_1)}{(M_{d_2}^2-\lm)}
\Bigg)\nn\\
&&\times \left(1-\dl_{d_1d_2}\frac{M_{d_2}^2}{\lm}\right)+ \mc O\left(\al^5, \bt^5\right)\Bigg]
+(M_Q\leftrightarrow M_D, \al \leftrightarrow \bt)\Bigg\}\nn\\
&+&\Bigg\{\sum_{Q_1<Q_2<Q_3<Q_4} 2\left(\prod_{i=1,4}\frac{ M_{Q_i}}{(M_{Q_i}^2-\lm)}\right)\nn\\
&&\times\Bigg[\sum_{d_1,d_2}(-\lm)^2\Bigg(
\frac{A_2(Q_1,Q_2;d_1,d_3 )A_2(Q_3,Q_4;d_1,d_3)}{(M_{d_1}^2-\lm)(M_{d_3}^2-\lm)}\nn\\
&&\times
\frac{B_2(d_2,d_4;Q_1,Q_2)B_2(d_2,d_4;Q_3,Q_4)}{(M_{d_2}^2-\lm)(M_{d_4}^2-\lm)}
+\frac{A_2(Q_1,Q_3;d_1,d_3 )A_2(Q_2,Q_4;d_1,d_3)}{(M_{d_1}^2-\lm)(M_{d_3}^2-\lm)}\nn\\
&&\times
\frac{B_2(d_2,d_4;Q_1,Q_3)B_2(d_2,d_4;Q_2,Q_4)}{(M_{d_2}^2-\lm)(M_{d_4}^2-\lm)}
+\frac{A_2(Q_1,Q_4;d_1,d_3 )A_2(Q_2,Q_3;d_1,d_3)}{(M_{d_1}^2-\lm)(M_{d_3}^2-\lm)}\nn\\
&&\times
\frac{B_2(d_2,d_4;Q_1,Q_4)B_2(d_2,d_4;Q_2,Q_3)}{(M_{d_2}^2-\lm)(M_{d_4}^2-\lm)}\Bigg)\nn\\
&&\times \left(1-\dl_{d_1d_2}\frac{M_{d_2}^2}{\lm}\right)
\left(1-\dl_{d_1d_4}\frac{M_{d_4}^2}{\lm}\right)
\left(1-\dl_{d_3d_2}\frac{M_{d_2}^2}{\lm}\right)
\left(1-\dl_{d_3d_4}\frac{M_{d_4}^2}{\lm}\right)+ \mc O\left(\al^6, \bt^6\right)\Bigg]\nn\\
&&+(M_Q\leftrightarrow M_D, \al \leftrightarrow \bt)\Bigg\} \ , \label{eq:chpolP1}
\ea
where all the discrete sums are taken from zero to infinity and $\lambda$ represents the eigenvalues of the $\left[M^\dag M\right]$ matrix.



\end{document}